\documentclass[11pt]{article}
\usepackage{amsmath,amssymb,color,epsfig,cite}
\usepackage{graphicx}
\usepackage{subfigure}
\usepackage{setspace}

\usepackage{amsfonts}

\newcommand{\hoch}[1]{$\, ^{#1}$}

\makeatletter
\@addtoreset{equation}{section}
\makeatother

\newcommand{\be}{\begin{equation}}
\newcommand{\ee}{\end{equation}}
\newcommand{\bea}{\setlength\arraycolsep{2pt} \begin{eqnarray}}
\newcommand{\eea}{\end{eqnarray}}
\newcommand{\nn}{\nonumber}

\def\ft#1#2{{\textstyle{\frac{\scriptstyle #1}{\scriptstyle #2} } }}
\def\fft#1#2{{\frac{#1}{#2}}}

\def\0{{\sst{(0)}}}
\def\1{{\sst{(1)}}}
\def\2{{\sst{(2)}}}
\def\3{{\sst{(3)}}}
\def\4{{\sst{(4)}}}
\def\5{{\sst{(5)}}}
\def\6{{\sst{(6)}}}
\def\7{{\sst{(7)}}}
\def\8{{\sst{(8)}}}
\def\9{{\sst{(9)}}}

\def\sst#1{{\scriptscriptstyle #1}}

\allowdisplaybreaks

\begin{document}



\begin{center}
{\large {\bf Holographic Aspects of Quasi-topological Gravity}}

\vspace{10pt}
Jun Peng\hoch{1\dagger} and Xing-Hui Feng\hoch{2*}

\vspace{15pt}

\hoch{1}{\it Department of Physics, Beijing Normal University, Beijing 100875, China}

\vspace{10pt}

\hoch{2}{\it Department of Physics, Tianjin University, Tianjin 300072, China}

\vspace{30pt}

\underline{ABSTRACT}

\end{center}

In this paper, we study the holography of quasi-topological gravity in several aspects. We redo the calculation of shear viscosity on the boundary CFT with a new method which is associated with conserved Noether current and show that it has only one mode explicitly. Then we study the butterfly effect in AdS planar black hole and find it has two butterfly velocity modes because of the quasi-topological term. We obtain new black hole solution through introducing matter fields. We calculate the thermoelectric DC conductivity with momentum dissipation in quasi-topological gravity and find its results are the same with those of Einstein and Gauss-Bonnet gravities. These results show us explicit similarities and differences between quasi-topological term and actual topological term in the context of holography.

\vfill {\footnotesize \hoch{\dagger} jun\_peng@mail.bnu.edu.cn \ \ \hoch{*}xhfeng@tju.edu.cn}

\pagebreak

\addtocontents{toc}{\protect\setcounter{tocdepth}{2}}


\newpage

\section{Introduction}

Holography which links a $(d + 1)$ dimensional gravity theory in the bulk to a dual $d$ dimensional quantum field theory on the boundary plays a central role in modern theoretical physics \cite{Maldacena:1997re,Gubser:1998bc,Witten:1998qj}. This duality not only provides a powerful tool to study strongly coupled field theories on the boundaries of some gravitational backgrounds, but also sheds light on the quantum aspects of gravity. Many great progress have been made in the framework of Einstein gravity. To match more complicated and realistic condensed matter phenomenons, we may take higher curvature terms into consideration. To some extend, this is also the consistent requirement of quantum gravity.

Higher curvature interactions are generically expected to arise for the UV completion of general relativity. It was realized more than forty years ago that the inclusion of quadratic terms in the gravitational action can lead to a power counting renormalizable theory of gravity \cite{Stelle:1976gc,Stelle:1977ry}. The theory admits the usual Schwarzschild black hole as a vacuum solution.  Recently it was demonstrated numerically that the theory contains a new black hole associated with the condensation of the massive spin-2 modes \cite{Lu:2015cqa,Lu:2015psa}.  However, due to the presence of potential massive ghost in the spin-2 modes, the unitarity in quantum theory will break down \cite{Sisman:2011gz}.

Gauss-Bonnet gravity is a particular combination of quadratic curvature terms, of which the equations of motion remain two derivatives with respect to arbitrary metric. As a result, it can avoids the ghost problem. A natural cubic generalization is third order Lovelock action, but it is trival in five and higher dimensions. Recently, a new cubic curvature interaction, named quasi-topological gravity, has been constructed in \cite{Oliva:2010eb,Myers:2010ru}. Quasi-topological gravity has two remarkable properties. First, when evaluated on spherically symmetric backgrounds, the equations of motion are only two derivatives and admit exact solutions whose form is very similar to Lovelock gravity. Second, the linearized equations of motion on maximally symmetric backgrounds coincide with the linearized Einstein equations up to an overall factor. Thus it is also free of massive spin-2 ghost for general higher curvature gravity.\cite{Myers:2010jv,Myers:2010tj}. The generalizations of quasi-topological gravity can be found in \cite{Dehghani:2011vu,Dehghani:2013ldu,Bueno:2016xff,Bueno:2016ypa,Cisterna:2017umf,Hennigar:2017ego,Ahmed:2017jod,Li:2017ncu,Li:2017txk}. It deserves to point out that one kind of infinite derivative gravities which are singularity and ghost free were constructed in \cite{Biswas:2011ar,Biswas:2013cha,Biswas:2016egy}.

Based on AdS/CFT correspondence, the presence of higher order derivatives terms in gravity means new couplings among operators in the dual CFT. So various higher curvature couplings will lead to more classes of dual field theories. Because the cubic quasi-topological term is non trival in five dimensions, we would get a more general four dimensional dual CFT. In this paper, we aim to focus on the holographic aspects of quasi-topological gravity in five dimension, including hydrodynamics, butterfly effect, thermoelectric DC conductivities.

This paper is organized as follows. In section 2, we give a brief review of quasi-topological gravity. In section 3, we calculate the shear viscosity on the boundary CFT with conserved Noether current and our result confirms the previous result using pole method \cite{Myers:2010jv}. In section 4, we study the holographic butterfly effect and obtain the butterfly velocities. In section 5, we calculate the thermoelectric DC conductivities with momentum dissipation through introducing spatial scalar fields. We conclude in section 6.

\section{Review of quasi-topological gravity}

We start with a brief review of five dimensional quasi-topological gravity. The action can be written as \cite{Myers:2010ru}:
\be
I = \frac{1}{16\pi}\int d^5x\sqrt{-g}\left[\frac{12}{\ell_0^2}+R+\frac{\lambda}{2}\ell_0^2{\cal X}_4+\frac{7}{8}\mu \ell_0^4{\cal Z}_5\right]
\ee
where ${\cal X}_4$ is Gauss-Bonnet term
\be
{\cal X}_4 = R_{\mu\nu\rho\sigma}R^{\mu\nu\rho\sigma}-4R_{\mu\nu}R^{\mu\nu}+R^2,
\ee
and ${\cal Z}_5$ is the cubic quasi-topological term
\bea
{\cal Z}_5 &=& R_a{}^c{}_b{}^d R_c{}^e{}_d{}^f R_e{}^a{}_f{}^b+\frac{1}{56}(21R_{abcd}R^{abcd}R-72R_{abcd}R^{abc}{}_e R^{de}\cr
&&+120R_{abcd}R^{ac}R^{bd}+144R_a{}^b R_b{}^c R_c{}^a-132R_a{}^b R_b{}^a R+15R^3).
\eea
The covariant equation of motion associated with the variation of the metric is
\be
{\cal E}_{ab} \equiv  P_{acde}R_b{}^{cde}-\frac{1}{2}g_{ab}{L}-2\nabla^c\nabla^dP_{acdb} = 0\,,
\ee
where $P^{\mu\nu\rho\sigma}$ is defined by $P^{\mu\nu\rho\sigma} = \frac{\partial L}{\partial R_{\mu\nu\rho\sigma}}$, whose explicit expression is given by appendix \cite{Lan:2017xcl}.

This theory admits an AdS vacua with radii
\be
\frac{1}{\ell^2} = \frac{f_\infty}{\ell_0^2}\label{adsradii}
\ee
where the constant $f_\infty$ is determined as one of the roots of
\be
1-f_\infty+\lambda f_\infty^2+\mu f_\infty^3 = 0.
\ee
The solutions describing planer AdS black holes take the form \cite{Myers:2010ru}
\be
ds^2 = \frac{r^2}{\ell_0^2}\left(-\frac{f(r)}{f_\infty}dt^2+dx_1^2+dx_2^2+dx_3^2\right)+\frac{\ell_0^2}{r^2f(r)}dr^2,
\ee
where $f(r)$ is determined by roots of the following cubic equation:
\be
1-f(r)+\lambda f(r)^2+\mu f(r)^3 = \frac{r_0^4}{r^4}.
\ee
It is easy to see that the black hole horizon is located at $r=r_0$ where $f(r_0)=0$.


\section{Holographic shear viscosity}

In this section, we compute the ratio of the shear viscosity to entropy density for five dimensional
quasi-topological gravity. By now, the holographic calculation of the shear viscosity for various models
is well understood, seeing \cite{Policastro:2001yc,KSS,KSS0,Buchel:2003tz,Buchel:2004qq,Benincasa:2006fu,Landsteiner:2007bd,Cremonini:2011iq,Iqbal:2008by,Cai:2008ph,Cai:2009zv,Brustein:2007jj,Liu:2015tqa,Sadeghi:2015vaa,horndeski1,horndeski2,Liu:2016way,Hartnoll:2016tri,Alberte:2016xja,Liu:2016njg,Parvizi:2017boc} for references. In fact the shear viscosity of quasi-topological gravity has already been investigated in \cite{Myers:2010jv} using pole method. Here we adopt an more universal and elegant formula proposed in \cite{Liu:2017kml}, which is especially convenient for high derivative gravities. This new method has been well developed in \cite{Fan:2018qnt}.

To study the holographic shear viscosity, we perform the transverse and traceless perturbation
\be
dx^2 \rightarrow dx^2+2\Psi(t,r)dxdy
\ee
We make an ansatz
\be
\Psi(t,r) = \zeta t+\psi(r)
\ee
where $\zeta$ is a constant. The linearized equation for perturbation $\psi(r)$ becomes
\be
A\psi''+B\psi'+C\psi = 0,\label{viscosityeq}
\ee
where the coefficients $A,B,C$ are respectively
\bea
A &=& r^2 f \left[9 \mu  r^4 f''^2+72 \mu  r^2 f'^2+f' \left(9 \mu  r^4 f^{(3)}+81 \mu  r^3 f''+8 \lambda  r\right)-8\right]\nn\\
&&+8 r^2 f^2 \left(3 \mu  r f'+2 \lambda \right)+24 \mu  r^2 f^3\nn\\
B &=& r^2 f' \left[9 \mu  r^4 f''^2+72 \mu  r^2 f'^2+f' \left(9 \mu  r^4 f^{(3)}+81 \mu  r^3 f''+8 \lambda  r\right)-8\right]\nn\\
&&+r f \left[162 \mu  r^4 f''^2+552 \mu  r^2 f'^2+f'' \left(27 \mu  r^5 f^{(3)}+8 \lambda  r^2\right)\right.\nn\\
&&\qquad\left.+f' \left(9 \mu  r^5 f^{(4)}+162 \mu  r^4 f^{(3)}+792 \mu  r^3 f''+80 \lambda  r\right)-40\right]\nn\\
&&+8 r f^2 \left(3 \mu  r^2 f''+27 \mu  r f'+10 \lambda \right)+120 \mu  r f^3\nn\\
C &=& -8[3 f^2 \left(\mu  r^2 f''+8 \mu  r f'+4 \lambda \right)+2 f \left(\lambda  r^2 f''+3 \mu  r^2 f'^2+8 \lambda  r f'-6\right)\nn\\
&&\qquad-r^2 f''+2 \lambda  r^2 f'^2-8 r f'+12 \mu  f^3+12]
\eea
Note that the linear time dependent $\zeta$ term vanishes in virtue of background equation of motion. What is somewhat surprising is that the linearized equation involves only two derivatives, though itself is very complicated. This distinguishes other high derivative gravities remarkably.

A rather universal connection between Noether current and boundary stress tensor proposed in \cite{Liu:2017kml} provides us a simple way to calculate the transport coefficients of dual field theory on the boundary. According to Wald's procedure, we can construct the radially conserved current
\be
{\cal J}^{x} = \sqrt{-g}J^{rx}
\ee
where
\be
J^{\mu\nu} = 2P^{\mu\nu\rho\sigma}\nabla_\rho\xi_\sigma+4\xi_\rho\nabla_\sigma P^{\mu\nu\rho\sigma}
\ee
To get a conserved current associated with boundary stress tensor which results from the perturbation $\Psi(t,r)$, $\xi$ is chosen to be spacelike Killing vector $\partial_y$. For the static background, the current ${\cal J}^{x}$ vanishes identically; it gives non-trivial contribution at the linear order once the perturbation $\Psi(t,r)$ is turned on. We find
\bea
16\pi{\cal J}^{x} &=& \frac{r^5 f \psi '}{8 \ell_0^5 \sqrt{f_\infty}} \left[9 \mu  r^4 f''^2+72 \mu  r^2 f'^2+f' \left(9 \mu  r^4 f^{(3)}+81 \mu  r^3 f''+8 \lambda  r\right)\right.\nn\\
&&\qquad\qquad+8 f \left(3 \mu  r f'+2 \lambda \right)+24 \mu  f^2-8\Big]
\eea
It can then be easily verified that the radial conservation law $\partial_r{\cal J}^x$ gives precisely the linearized perturbation equation \eqref{viscosityeq}.

Since the linearized perturbation equation is still two derivatives, we can impose the ingoing horizon boundary conditions for $\psi$ as
\be
\psi = \zeta\frac{\log(r-r_0)}{4\pi T}+\cdots
\ee
where $T$ is Hawking temperature, which can be easily calculated
\be
T=\frac{r_0}{\pi\ell_0^2}\frac{1}{f_\infty^{1/2}}
\ee
Substituting this horizon data into the conserved current, we have
\be
16\pi{\cal J}^x = \frac{r_0^3}{\ell_0^3}[1-4\lambda-36\mu(9-64\lambda+128\lambda^2+48\mu)]\zeta
\ee
So the viscosity is given by
\be
\eta = \frac{\partial{\cal J}^x}{\partial\zeta} = \frac{r_0^3}{16\pi\ell_0^3}[1-4\lambda-36\mu(9-64\lambda+128\lambda^2+48\mu)]
\ee
The entropy density can be calculated using Wald entropy formula
\be
s=\frac{r_0^3}{4\ell_0^3}
\ee
The viscosity/entropy ratio is
\be
\frac{\eta}{s} = \frac{1}{4\pi}[1-4\lambda-36\mu(9-64\lambda+128\lambda^2+48\mu)]
\ee
This result has already been obtained in \cite{Myers:2010jv} using pole method. Our calculation confirmed this in a rigorous way. To be specific, we have shown explicitly the linearized perturbation equation only involves two derivative, so this is the unique mode of viscosity. This property resembles Gauss-Bonnet gravity instead of general high derivative gravities.

\section{Holographic butterfly effect}
\label{sec:butterfly}
The butterfly effect is associated with the exponential growth of a small perturbation to a
quantum system. In the context of holography, this effect has a beautiful realization \cite{Shenker:2013pqa,Shenker:2013yza,Roberts:2014isa,Shenker:2014cwa,Maldacena:2015waa} in
terms of a gravitational shock wave near the horizon of an AdS black hole \cite{Sfetsos:1994xa}.
The butterfly velocities for a variety of AdS planar black holes of various matter energy momentum
tensor were obtained \cite{Blake:2016wvh,Roberts:2016wdl,Feng:2017wvc}. The study has been further generalized to include
higher-order gravities \cite{Roberts:2014isa,Alishahiha:2016cjk,Qaemmaqami:2017bdn,Qaemmaqami:2017jxz,Huang:2017ohr}. The expression of $v_B$
can be simple or complicated depending
on the detail structures of the black holes.

We study the butterfly effect of AdS planar black hole with the metric
\be
ds^2 = -f(r)dt^2+\frac{dr^2}{f(r)}+\frac{r^2}{\ell^2}dx^idx^i\label{butterflymetric}
\ee
With this ansatz, the solution is determined by
\be
1-\frac{\ell_0^2f}{r^2}+\lambda\frac{\ell_0^4f^2}{r^4}+\mu\frac{\ell_0^6f^3}{r^6} = \frac{r_0^4}{r^4}
\ee
The temperature of black hole is
\be
T = \frac{f'(r_0)}{4\pi} = \frac{r_0}{\ell_0^2\pi}
\ee
In the asymptotic infinity, $f(r)$ behaves as
\be
f(r) \sim \frac{r^2}{\ell^2}
\ee
Near the horizon $r=r_0$, the function $f$ can be expressed as
\be
f=f_1 (r-r_0) + f_2 (r-r_0)^2 + \cdots\,.\label{ftaylor}
\ee
To study the butterfly effects, we need convert to the Kruskal coordinates $(u,v)$
\be
u=e^{\kappa (r_* - t)}\,,\qquad v=- e^{\kappa (r_*+t)}\,,\qquad
\hbox{with}\qquad dr_* = \fft{dr}{f}\,.
\ee
Here $\kappa=2\pi T=\ft12f_1$ is the surface gravity on the horizon $r=r_0$, which corresponds to $uv=0$.  Near the horizon, we have
\be
u v = (r-r_0) - \fft{f_2}{f_1} (r-r_0)^2 + \cdots\,,\qquad
r-r_0 = uv + \fft{f_2}{f_1}\, (uv)^2 + \cdots\,.
\ee
The metric (\ref{butterflymetric}) can now be expressed as
\be
ds_5^2 = A(u v) du dv + B(uv) d x^i d x^i\,,
\ee
where
\be
A(u v) = \fft{1}{\kappa^2} \fft{f}{uv}\,,\qquad B(uv) = \frac{r^2}{\ell^2}\,.
\ee
We can Taylor expand the functions $A$ and $B$ on the horizon $uv=0$,
\be
A=A_0 + A_1\, (uv) + A_2\, (uv)^2 + \cdots\,,\qquad
B=B_0 + B_1\, (uv) + B_2\, (uv)^2 + \cdots\,.\label{AB}
\ee
The relation between the coefficients $(A_i,B_i)$ and $f_i$ in (\ref{ftaylor}) can be found in \cite{Feng:2017wvc}. To calculate the butterfly velocity, one considers the metric perturbation
\be
ds^2 = A(uv)\,dudv + B(uv)\, dx^i dx^i - A(uv)\,\delta(u)\,h(\vec x)\,du^2.
\ee
We find that the linear butterfly equation of motion for $h(\vec x)$ is
\be
\frac{12A_0(8\lambda-3)\mu}{B_1}\Box\Box h(\vec x)+\Box h(\vec x)-\frac{3B_1}{A_0}h(\vec x) \sim E e^{2\pi T\, t_w} \delta(\vec x)  \,.\label{butterflyeom}
\ee
When $\mu=0$, this is exactly the shockwave equation for Einstein and Gauss-Bonnet theories. For general $\mu$, the shock equation can be factorized as
\be
(\Box+m_1^2)(\Box+m_2^2)h(\vec x) \sim E e^{2\pi T\, t_w} \delta(\vec x).\label{butterflyeom2}
\ee
Here $\Box$ is the Laplacian of the flat space $dx^idx^i$. There are two modes with masses
\bea
m_1^2 &=& \frac{6B_1}{A_0(1-\sqrt{1+144(8\lambda-3)\mu})}\nn\\
m_2^2 &=& \frac{6B_1}{A_0(1+\sqrt{1+144(8\lambda-3)\mu})}
\eea
Stability requires that both masses squared are non-negative, which leads to a new constraint about coupling constants of the high derivative terms (see casuality constraints in \cite{Myers:2010jv})
\be
-\frac{1}{144}\le(8\lambda-3)\mu\le0
\ee
The solution to \eqref{butterflyeom2} takes the form
\be
h(\vec x) \sim \frac{E/(m_1^2-m_2^2)}{\sqrt{|\vec x|}}\left(e^{2\pi T(t_w-t_\ast)-m_1|\vec x|}-e^{2\pi T(t_w-t_\ast)-m_2|\vec x|}\right)
\ee
The butterfly velocity is defined as
\be
v_B = \frac{2\pi T}{m}.
\ee
Thus the corresponding butterfly velocities associated with the $m_1$ and $m_2$ modes are given by respectively
\bea
v_1^2 &=& \frac{1}{3f_\infty}(1-\sqrt{1+144(8\lambda-3)\mu})\nn\\
v_2^2 &=& \frac{1}{3f_\infty}(1+\sqrt{1+144(8\lambda-3)\mu})
\eea
When we set $\mu=0$, the $v_1$ vanishes and $v_2=\frac{2}{3f_\infty}$, which is just the result of Gauss-Bonnet gravity.

\section{Thermoelectric DC conductivities with momentum dissipation}

A small electric field $E$ and thermal gradient $\nabla T$
will generate an electric current $J$ and thermal
current $Q$. The conductivity matrix is defined through
\begin{equation}
\left(
  \begin{array}{c}
    J \\
    Q \\
  \end{array}
\right)
=
\left(
  \begin{array}{cc}
    \sigma  & \alpha T  \\
    \bar \alpha T  & \kappa T \\
  \end{array}
  \right)
  \left(
  \begin{array}{c}
    E \\
    - (\nabla T)/T \\
  \end{array}
\right)           \,,
\end{equation}
where $\sigma$ is the electric conductivity, $\bar \kappa$ is the thermal
conductivity and $\alpha\,,\bar \alpha $ are thermoelectric conductivities.
In \cite{Donos:2014cya}, Donos and Gauntlett gave a strategy to calculate holographic DC conductivities with momentum dissipation through introducing spatial axion fields. In this approach, we need only to focus on the horizon data by analysing the boundary condition on the horizon. The key step is to construct the heat current $Q$ which is difficult for high derivative gravities. Thanks to the relationship between the heat current and Noether current built in \cite{Liu:2017kml}, this process becomes rather systematic. Further discussions on the holographic DC conductivities in various theories can be found in \cite{Amoretti:2014mma,lat6,gsge,DC4,DCsera,Donos:2017oym,Jiang:2017imk,Liu:2017kml,Mokhtari:2017vyz,Li:2017nxh}. The holographic s-wave and p-wave superconductor of quasi-topological gravity have been investigated in \cite{Kuang:2010jc,Kuang:2011dy}

Firstly we need to construct charged black brane with axions through introducing a Maxwell field and three axion fields into the action
\be
S = \frac{1}{16\pi G}\int d^5x\sqrt{-g}\left(R+\frac{12}{\ell_0^2}+\frac{\lambda}{2}\ell_0^2{\cal X}_4+\frac{7}{8}\mu \ell_0^4{\cal Z}_5-\frac{1}{2}\sum^3_{i=1}(\partial\chi_i)^2-\frac{1}{4}F^2\right)
\ee
where $\chi_i(i=1,2,3)$ are three axion fields.
We choose the following planar symmetric ansatz
\bea
ds^2 &=& -fdt^2+\frac{dr^2}{f}+r^2dx^idx^i\nn\\
A &=& a dt,\quad \chi_i=\beta x_i
\eea
The solution of Maxwell equation is
\be
a = \delta\left(1-\frac{r_0^2}{r^2}\right)
\ee
which vanishes on the horizon at $r=r_0$, and the constant $\delta$ can be interpreted as the chemical potential of the boundary CFT. This theory admit the black hole solution determined by the following algebraic relation
\be
\frac{\ell_0^2\delta^2r_0^4}{3r^6}-\frac{\ell_0^2\beta^2}{4r^2}+1-\frac{\ell_0^2f}{r^2}+\lambda\frac{\ell_0^4f^2}{r^4}+\mu\frac{\ell_0^6f^3}{r^6} = \frac{m}{r^4}
\ee
One can solve for the mass parameter $m$ from $f(r_0)=0$,
\be
m = \frac{\ell_0^2\delta^2r_0^2}{3}-\frac{\ell_0^2\beta^2r_0^2}{4}+r_0^4
\ee
The Hawking temperature is
\be
T = \frac{24r_0-3\ell_0^2\beta^2-4\ell^2\delta^2}{24\ell_0^2\pi r_0}\label{temperature}
\ee
In the asymptotic infinity, $f(r)$ behaves as
\be
f(r) \sim \frac{r^2}{\ell^2}
\ee
Then we can calculate the thermoelectric DC conductivities in holographic way. We consider the linear perturbation
\bea
ds^2 &=& - f dt^2 + \fft{dr^2}{f} + r^2 dx^i dx^i + 2 \Psi_1 dt dx_1 \,,\cr
A &=& a dt + \Psi_2 dx_1 \,, \quad \chi_1 = \beta x_1 + \Psi_3 \,,
\eea
with
\be
\Psi_1 = - t \zeta f + \psi_1(r) \,, \quad \Psi_2 = (- E + \zeta  a) t + \psi_2(r) \,,\qquad
\Psi_3=\psi_3(r)\,.
\ee
Put the above perturbation into the equations of motion, we can obtain tree linearized equations for $(\psi_1,\psi_2,\psi_3)$.
\bea
&&2 \ell_0^2 r^2 f \Big(3 \mu  \ell_0^4 f^2+2 \lambda  \ell_0^2 r^2 f-r^4\Big)\psi_1''-2 \ell_0^2 r f \Big(r^3 \left(r-2 \lambda  \ell_0^2 f'\right)+2 \ell_0^2 r f \left(\lambda  r-3 \ell_0^2 \mu  f'\right)+9 \mu  \ell_0^4 f^2\Big)\psi_1'\nn\\
&&- \Big[r^4 \Big(\ell_0^2 r^2 a'^2-2 \ell_0^2 r^2 f''+4 \lambda  \ell_0^4 f'^2-8 \ell_0^2 r f'-3 \beta ^2 \ell_0^2+24 r^2\Big)+6 \mu  \ell_0^6 r^2 f^2 f''\nn\\
&&\quad+4 \ell_0^2 r^2 f \Big(r^2 \left(\lambda  \ell_0^2 f''-3\right)+3 \mu  \ell_0^4 f'^2+2 \lambda  \ell_0^2 r f'\Big)-12 \mu  \ell_0^6 f^3\Big]\psi_1-2 \ell_0^2 r^6 f a' \psi_2'=0\\
&&\Big[r(\psi_1a'+f\psi_2')\Big]'=0,\qquad\psi_3''+\Big(\frac{f'}{f}+\frac{3}{r}\Big)\psi_3'-\frac{\beta\zeta}{r^2f}=0
\eea
Once again the linearized equation for $\psi_1$ which results from the generalized Einstein equation is still two derivatives.

The conserved electric current can be easily defined from the Maxwell equation and is given by
\be
J = \sqrt g F^{rx_1} =r(\psi_1 a' + f \psi_2')\,.
\ee
The holographic heat current can be constructed by using the method proposed in \cite{Liu:2017kml}, namely
\bea
Q &=& \sqrt{-g} (  2 P^{rx_1\rho\sigma} \nabla_\rho\xi_\sigma + 4 \xi_\rho \nabla_\sigma P^{rx_1\rho\sigma} + a F^{rx_1} )\nn\\
&=& \frac{1}{r^3}[r^4 a \left(\psi_1 a'+f \psi_2'\right)+\left(f \psi_1'-\psi_1 f'\right) \left(-3 \ell_0^4 \mu  f^2-2 \ell_0^2 \lambda  r^2 f+r^4\right)] \,.\label{heatcurrent}
\eea
Note that here $\xi$ is chosen to be the timelike Killing vector $\partial_t $. It can be checked that the linearized equations for $(\psi_1,\psi_2)$ is corresponding to $J'=0$ and $Q'=0$. Thus $J$ and $Q$ are indeed radially conserved. Since the linearized perturbation equations are two derivatives, we can still choose the ingoing boundary condition on the horizon as
\bea
\Psi_1 &\sim& \psi_{10}-\zeta f\frac{\log(r-r_0)}{4\pi T}+{\cal O}(r-r_0)+\cdots,\nn\\
\Psi_2 &\sim& \psi_{20}+(-E+\zeta a)\frac{\log(r-r_0)}{4\pi T}+{\cal O}(r-r_0)+\cdots,
\eea
where $\psi_{10}$ and $\psi_{20}$ are constants. In fact, $\psi_{20}$ is a pure gauge and $\psi_{10}$ can be determined through the equation of motion
\be
\psi_{10} = -\frac{2Er_0\delta+\zeta r_0^2f'(r_0)}{\beta^2}.
\ee
The electric and heat currents, evaluated on the horizon, are thus given by
\bea
J &=& \left(r_0+\frac{4\delta^2r_0}{\beta^2}\right)E+\frac{8\pi \delta r_0^2T}{\beta^2}\zeta\nn\\
Q &=& \frac{8\pi \delta r_0^2T}{\beta^2}E+\frac{16\pi^2r_0^3T^2}{\beta^2}\zeta
\eea
The elements of thermoelectric DC conductivity matrix can now be obtained
\bea
\sigma &=& \fft{\partial J}{\partial E} = r_0 + \frac{4\delta^2r_0}{\beta ^2} \,, \qquad
\alpha = \fft 1 T \fft{\partial J}{\partial \zeta}= \frac{8 \pi  \delta r_0^2}{\beta ^2} \,, \cr
\bar \alpha &=& \fft 1 T \fft{\partial Q }{\partial E}= \frac{8 \pi  \delta r_0^2}{\beta ^2} \,,\qquad
\bar \kappa = \fft 1 T \fft{\partial Q}{\partial \zeta } = \frac{16 \pi ^2 r_0^3 T }{\beta ^2} \,.
\label{dcmatrix}
\eea
Interestingly, these results coincide with those of Einstein \cite{Donos:2014cya} and Gauss-Bonnet \cite{gsge} theory. In other words, Gauss-Bonnet term and quasi-topological term don't modify the thermoelectric DC conductivities. This can be seen from the heat current \eqref{heatcurrent}. Though high derivative terms contribute to the expression of heat current $Q$, the value doesn't change when evaluated on the horizon, because the horizon is determined by $f=0$.

\section{Conclusion}

Quasi-topological gravity has many well properties. Firstly its linearized equation around maximally symmetric vacuum is identical to the one of Einstein and Gauss-Bonnet gravities up to certain overall factor. Its static black hole solution is determined by an algebraic equation which is similar to Gauss-Bonnet gravity. In this paper, we studied the holographic application of these black holes. We calculated the shear viscosity on the boundary CFT with a new method and confirmed the result with previous literature. This new approach is universal and especially convenient for high derivative gravities. We also have shown that it is the unique mode in this special high derivative gravity, i.e. the linearized perturbation equation involves only two derivatives, which contrasts with that for general high derivative gravities. Then we studied the butterfly effect of quasi-topological gravity. To obtain a shock wave equation associated with butterfly effect for quasi-topological gravity, in fact we need perform a totally different perturbation around static AdS planar black hole, which leads to a fourth order linearized perturbation equation. As a result, we obtained two butterfly velocity modes. This is very different from Gauss-bonnet gravity. After introducing matter fields such as Maxwell field and axion fields, we obtained new black hole solution. We calculated the thermoelectric DC conductivities with momentum dissipation in this black hole background. We found that the linearized perturbation equations still involve only two derivatives and furthermore the results are the same with those of Einstein and Gauss-Bonnet gravities. Our results give more interesting properties of quasi-topological gravity. Its physical implication deserves further research.

\section*{Acknowledgement}

We are grateful to Prof. Hong L\"u for useful assistance. This work is supported in part by NSFC grants No.~11475024, No.~11175269 and No.~11235003.

\appendix

\section{Explicit expression of tensor $P^{\mu\nu\rho\sigma}$}

The $P^{\mu\nu\rho\sigma}$ for Ricci scalar $R$ is
\be
P_1^{\mu\nu\rho\sigma} = \frac{1}{2}(g^{\mu\rho}g^{\nu\sigma}-g^{\mu\sigma}g^{\nu\rho}),
\ee
The $P^{\mu\nu\rho\sigma}$ for Gauss-Bonnet term ${\cal X}_4$ is
\bea
P_2^{\mu\nu\rho\sigma} &=& 2R^{\mu\nu\rho\sigma}+R(g^{\mu\rho}g^{\nu\sigma}-g^{\mu\sigma}g^{\nu\rho})\nn\\
&&-2(g^{\mu\rho}R^{\nu\sigma}-g^{\mu\sigma}R^{\nu\rho}-g^{\nu\rho}R^{\mu\sigma}+g^{\nu\sigma}R^{\mu\rho})
\eea
The $P^{\mu\nu\rho\sigma}$ for quasi-topological term ${\cal Z}_5$ is
\begin{align}
P_3^{\mu\nu\rho\sigma}&=\frac{3}{4}(R^{\nu e\sigma f}R^\mu{}_e{}^\rho{}_f-R^{\nu e\rho f}R^\mu{}_e{}^\sigma{}_f-\mu\leftrightarrow\nu)\cr
&+\frac{1}{56}\bigg(21[2R^{\mu\nu\rho\sigma} R+\frac{1}{2}R_{abcd}R^{abcd}(g^{\mu\rho}g^{\nu\sigma}-g^{\mu\sigma}g^{\nu\rho})]\cr
&-18[2(R^{\mu\nu\rho}{}_e R^{\sigma e}-R^{\mu\nu\sigma}{}_e R^{\rho e}+R^{\rho\sigma\mu}{}_e R^{\nu e}-R^{\rho\sigma\nu}{}_e R^{\mu e})\cr
&+(R_{abc}{}^\nu R^{abc\sigma}g^{\mu\rho}-R_{abc}{}^\nu R^{abc\rho}g^{\mu\sigma}-\mu\leftrightarrow\nu)]\cr
&+60[(R^{\mu\rho}R^{\nu\sigma}-R^{\mu\sigma}R^{\nu\rho})+(R^\nu{}_b{}^\sigma{}_d R^{bd}g^{\mu\rho}-R^\nu{}_b{}^\rho{}_d R^{bd}g^{\mu\sigma}-
\mu\leftrightarrow\nu)]\cr
&+108[g^{\mu\rho}R^{\sigma c}R_c{}^\nu-g^{\mu\sigma}R^{\rho c}R_c{}^\nu-\mu\leftrightarrow\nu]\cr
&-66[(g^{\mu\rho}R^{\nu\sigma}R-g^{\mu\sigma}R^{\nu\rho}R-\mu\leftrightarrow\nu)+R_{ab}R^{ab}(g^{\mu\rho}g^{\nu\sigma}-g^{\mu\sigma}g^{\nu\rho})]\cr
&+\frac{45}{2}R^2(g^{\mu\rho}g^{\nu\sigma}-g^{\mu\sigma}g^{\nu\rho})\bigg)\label{generalP}
\end{align}

\end{document}